\documentclass[twocolumn, superscriptaddress,preprintnumbers,amsmath,amssymb, longbibliography]{revtex4-1}
\usepackage{graphicx}
\usepackage{dcolumn}
\usepackage{bm}
\usepackage{mathtools}
\usepackage{epstopdf}
\usepackage{float}

\begin{document}

\preprint{Version: \today}

\title{Tunable Superconducting Josephson Dielectric Metamaterial}

\author{M. Trepanier}

\affiliation{Center for Nanophysics and Advanced Materials, Department of Physics, University of Maryland, College Park, MD 20742-4111, USA}
\author{Daimeng Zhang}
\affiliation{Department of Electrical and Computer Engineering, University of Maryland, College Park, MD 20742-3285, USA}
\author{L. V. Filippenko}
\affiliation{Laboratory of Superconducting Devices for Signal Detection and Processing, Kotelnikov Institute of Radio Engineering and Electronics, Moscow 125009, Russia}
\author{V. P. Koshelets}
\affiliation{Laboratory of Superconducting Devices for Signal Detection and Processing, Kotelnikov Institute of Radio Engineering and Electronics, Moscow 125009, Russia}
\author{Steven M. Anlage}
\email{anlage@umd.edu}
\affiliation{Center for Nanophysics and Advanced Materials, Department of Physics,  University of Maryland, College Park, MD 20742-4111, USA}
\affiliation{Department of Electrical and Computer Engineering, University of Maryland, College Park, MD 20742-3285, USA}
\date{\today}
\begin{abstract}
We demonstrate a low-dissipation dielectric metamaterial with tunable properties based on the Josephson effect.  Superconducting wires loaded with regularly spaced Josephson junctions (critical current $I_c \approx 0.25$ $\mu$A) spanning a K-band waveguide and aligned with the microwave electric fields create a superconducting dielectric metamaterial.  Applied dc current tunes the cutoff frequency and effective permittivity of this unique electric metamaterial.  The results are in agreement with an analytical model for microwave transmission through the artificial dielectric medium.\end{abstract}
\maketitle

Metamaterials are artificially structured media with electromagnetic properties arising from the structure of individual meta-atoms and the interactions between them. Metamaterials have demonstrated emergent properties not seen in natural materials, e.g. a negative index of refraction \cite{Veselago1968, Smith2000, Shelby2001}, cloaking \cite{Alu2003, Schurig2006}, and super-resolution imaging \cite{Pendry2000, Jacob2006}.  Here we focus on wire-based dielectric metamaterials that interact primarily with oscillating electric fields and display values of the dielectric function that are rarely encountered in nature.  Such metamaterials have been studied in the past mainly for their anisotropic optical properties \cite{WireMMReview2012}, for creating deep sub-wavelength resonators \cite{Kolb2011}, or for their modification of the index of refraction with magnetic field \cite{FerriteWires2007, FerriteWires2014}.

The properties of conventional metamaterials are determined by their geometry and are typically fixed at the time of fabrication. Superconducting metamaterials \cite{Ricci2005, Anlage2011, Jung2014a, LazPhysRep2018} are tunable by means of the kinetic inductance of the superconducting electrons \cite{Ricci2007a}. The kinetic inductance can be increased by systematically suppressing superconductivity, which can be achieved by increasing temperature \cite{Meservey1969, KurterPlas2013} or by introducing currents \cite{Anlage1989} or magnetic fields. 
However, these tuning techniques are often limited in speed and tend to increase the dissipation, partially compromising the advantages of superconductivity.

Including Josephson junctions (a thin tunneling layer between two superconductors) allows the inductance of superconducting circuits to be tuned over a larger range and on faster time scales without a significant increase in dissipation. An ideal Josephson junction (JJ) has a variable inductance given by
\begin{equation}
\label{Ljj}
L_{JJ}=\frac{\Phi_0}{2 \pi I_c \cos{\delta}}
\end{equation}
where $\Phi_0=h/2e \cong 2.07 \times 10^{-15}$ Tm$^2$ is the flux quantum, $I_c$ is the critical current of the junction, and $\delta$ is the gauge-invariant phase difference of the superconducting order parameter across the junction. Under appropriate conditions, $L_{JJ}$ can controllably explore a variety of positive and negative values \cite{NegInduc1971, Rifkin1976}.

Josephson junctions have been used to create tunable and nonlinear magnetic metamaterials based mainly on radio frequency (rf) superconducting quantum interference devices (SQUIDs) \cite{Jung2013, Butz2013a, Trepanier2013, Zhang2015, DaimengIMD2016, TrepPRE2017, ZhuravelAPL2019}. Here we consider a complimentary electric metamaterial also tunable by means of the Josephson effect.  An earlier effort developed a planar Josephson electric metamaterial by capacitively coupling JJ-loaded wires to the center conductor of a co-planar waveguide \cite{Butz2014}.  The inductance was tuned by means of magnetic flux, as opposed to the present method of direct current bias through the junctions.

An array of thin wires spanning a single-conductor waveguide aligned with the polarization direction of the electric field of the propagating mode can introduce a cutoff frequency $\omega_c$ in the GHz range determined by the size and spacing of the wires \cite{Pendry1996, Walser2001, Ricci2007}. The wire array can be modeled by an effective dielectric function which resembles that of a lossless metal near its plasma edge, 
\begin{equation}
\label{epsr}
\epsilon_{r, eff}=1-F\left(\frac{\omega_c}{\omega}\right)^2
\end{equation}
where $\omega$ is frequency and $F$ is the filling fraction of the metamaterial in the waveguide.
We incorporate Josephson junctions into the wires to allow tuning of the cutoff frequency by means of current and temperature and thus tune the effective permittivity of the material.

Josephson junctions arrays have been extensively studied in 1D and 2D in an effort to make the junctions coherently emit microwaves in response to a dc voltage bias \cite{Tilley1970, Rogovin1976}. These studies have focused on dc bias dependence and microwave emission, generally in the absence of a uniform rf drive \cite{Jain1984, Benz1991, Barbara1999}. The current-biased 1D system can be mapped on to the Kuramoto model \cite{Acebron2005, Marvel2009} to understand the degree of coherent oscillation of the junctions. When a uniform rf drive is applied to an array of Josephson junctions there are giant Shapiro steps, which are used as a voltage standard \cite{Hamilton1990}. 2D Josephson junction arrays can also exhibit fractional giant Shapiro steps \cite{Benz1990}.

Here, in contrast, we are interested in how Josephson junction arrays behave as an electric metamaterial for tunable transmission of microwaves through the effective medium. We consider the metamaterial properties of a set of 1D arrays of wires each containing 100 junctions, in the classical limit.  To our knowledge this situation has not previously been experimentally measured or considered in detail theoretically.

The remainder of this paper is organized as follows: we begin with a description of how transmission through the metamaterial is numerically simulated and experimentally measured; we then show how the transmission and effective permittivity of a Josephson junction loaded wire metamaterial tunes in response to dc current and temperature in the low rf and dc current limits; we discuss the degree of agreement between the model and the data; and we close with conclusions.

\section{Modeling and Simulations}
The model of the JJ-loaded wire arrays begins with a single junction. 
Using the resistively and capacitively shunted junction (RCSJ) model \cite{Orlando1991}, the gauge-invariant phase difference across a single junction $\delta(t)$ driven by dc and rf current is given by
\begin{equation}
\frac{I_{dc}}{I_c}+\frac{I_{rf}}{I_c}\sin{\Omega \tau}=
\sin{\delta}+\gamma \frac{d \delta}{d \tau}+\frac{d^2 \delta}{d\tau^2}
\label{nlJJ}
\end{equation}
where $\tau=\omega_{p} t$, $\Omega=\omega/\omega_{p}$, $\omega_{p}=\sqrt{1/L_{JJ}(\delta=0)C}$ is the plasma frequency of the junction, and $\gamma=\frac{1}{R}\sqrt{\frac{L_{JJ}(\delta=0)}{C}}$ is the damping parameter. In this case the JJ is biased with both dc current ($I_{dc}$) and rf current ($I_{rf}$) at frequency $\omega$.  There are no chaotic solutions in the low rf current limit ($I_{rf}/I_c \ll 1$) considered in this paper \cite{Kautz1985}. 

To calculate the effective permittivity $\epsilon_r$ presented by the JJ-loaded wire arrays in the waveguide, the Josephson inductance is found (Eq. \ref{Ljj}) by using the time-averaged value of $\delta(t)$ (solution to Eq. \ref{nlJJ}) and this is used to find the cutoff frequency for a wire array spanning a single-conductor waveguide,
\begin{equation}
\omega_c=\frac{1}{d}\sqrt{\frac{l}{(L_{geo}+N L_{JJ})\epsilon_0}}
\end{equation}
where $d$ is the spacing between the wires, $l$ is the length of the wires, $L_{geo}$ is the geometric inductance of the superconducting wire, and $N$ is the number of junctions per wire \cite{Smith1999}. This model is only valid when Eq. \ref{Ljj} is a good approximation \textit{i.e.} low rf current $I_{rf}$ and $I_{dc}<I_c$. This model also assumes that all the junctions are identical, each 1D array is biased with the same current, and that they only interact with each other through their shared current.  Hence we expect the cutoff frequency of the metamaterial, and its dielectric properties, to be tuned through variation of $L_{JJ}$ by means of either dc current bias or temperature.

We treat the wires as an effective medium embedded in an empty waveguide. Transmission $S_{21}$ can be calculated by enforcing boundary conditions \textit{i.e.} $E$ and $H$ fields must be continuous at the boundaries of the effective medium and the empty waveguide \cite{TrepThesis2015}.

The length of the wires, the spacing between them, and their inductance \cite{Meservey1969} are determined by the lithographic process and have minimal variation. The critical current of the parallel array of junctions for a given temperature can be determined by measuring either the sudden change in microwave transmission, or onset of dc voltage, as a function of increasing dc current. The ratio of resistance and capacitance is determined by the re-trapping current, which is measured like $I_c$ but with decreasing dc bias current. For analytical modeling of the data, the length of the medium is set to be twice the spacing of the wires and the filling fraction (the only fitting parameter) is adjusted to match the magnitude of the observed tuning.

\begin{figure}[h]
\includegraphics*[width=80mm]{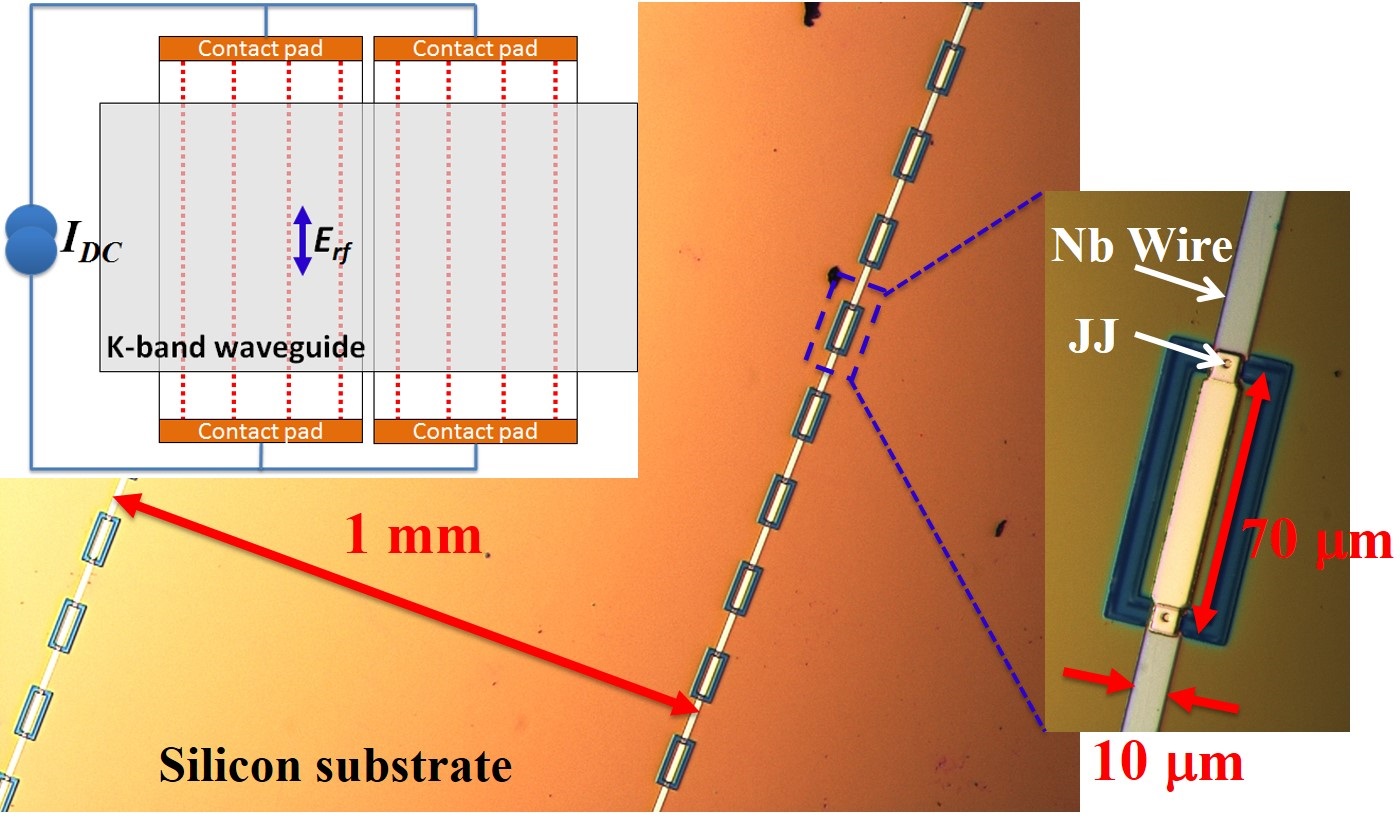}
\caption{Micrograph of Josephson junction loaded Nb wire arrays on Si substrate.  The wires are $10 \mu$m wide and the junctions punctuate the wires every $70 \mu$m.  A set of 4 wire arrays are spaced 1 mm apart on the chip.  Right inset shows detail of two JJs and connecting wire segments.  Left inset schematically shows the two chips loaded into the K-band waveguide and JJ arrays (dotted lines) biased by a dc current, and rf electric field from a passing electromagnetic wave in the waveguide.}
\label{NewWiresFig}
\end{figure}

\section{Experimental Setup}
\begin{figure}[h]
\includegraphics*[width=75mm]{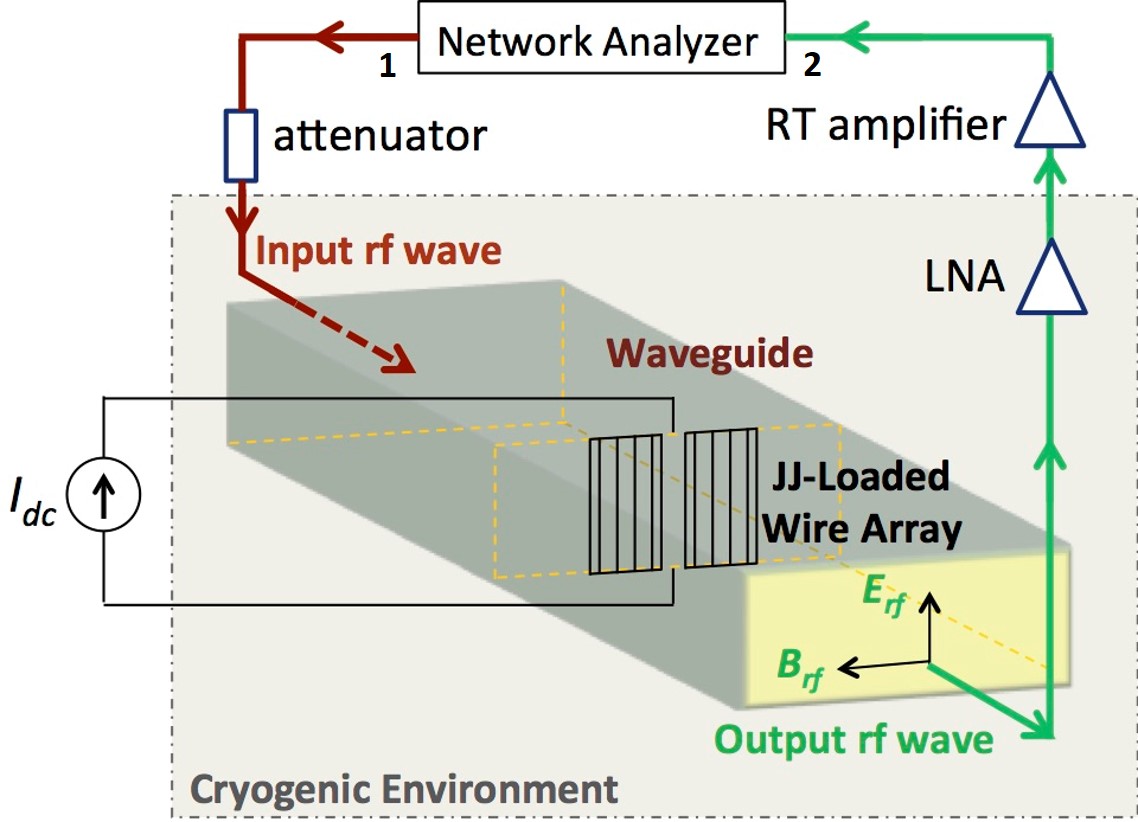}
\caption{Experimental setup schematic showing the flow of microwaves from the network analyzer through an attenuator, the waveguide and metamaterial, and then out through cryogenic and room temperature amplifiers before entering the network analyzer.  Also shown is the orientation of the JJ-loaded wire arrays and propagating rf electric field in the K-band waveguide.}
\label{fig1}
\end{figure}

The Nb-AlO$_x$-Nb JJ-loaded wire arrays were prepared at the Kotelnikov IREE \cite{Koshelets2003} and designed to have a cutoff frequency that tunes through the measurable frequency range of the single-mode propagation K-band waveguide, $14-22$ GHz. The Nb wires are 10 $\mu$m wide and consist of alternating segments that are either 135 nm or 300 nm thick (see Fig. \ref{NewWiresFig}).  There are four equally spaced (1 mm) wires of $N=100$ junctions each on a 4x8 mm$^2$ silicon chip. The junctions are spaced 70 $\mu$m apart and are nominally identical, with critical currents of $I_c(T=$ 4.5 K$)=0.25$ $\mu$A and the minimum amount of overlap capacitance. There are gold contact pads at each end of the wires to apply dc current bias; the contacts are normal metal so that the wires are not part of superconducting loops. We assume that each junction in the array is biased with the same dc current, thus acting coherently, and the results are consistent with this assumption.  Two chips (eight total wires) are oriented in a copper K-band rectangular waveguide so that the wires are parallel to the rf electric field as shown in Fig. \ref{NewWiresFig} (left inset) and Fig. \ref{fig1}. A single layer of wires is utilized. The transmission scattering parameter $S_{21}(\omega)$ through the waveguide is measured by means of a network analyzer at room temperature while the superconducting metamaterial and waveguide are maintained at a fixed temperature in a cryogenic environment (Fig. \ref{fig1}).  The microwave power reaching the wires is -50 dBm, inducing rf currents in the wires of 1 nA or less, satisfying the inequality $I_{rf}\ll I_c$.

\section{Results}
The cutoff frequency of the superconducting metamaterial wire array is tunable with temperature and dc bias current, allowing tuning of the effective permittivity and causing a measurable change in microwave transmission.

\begin{figure}[h]
\includegraphics*[width=75mm]{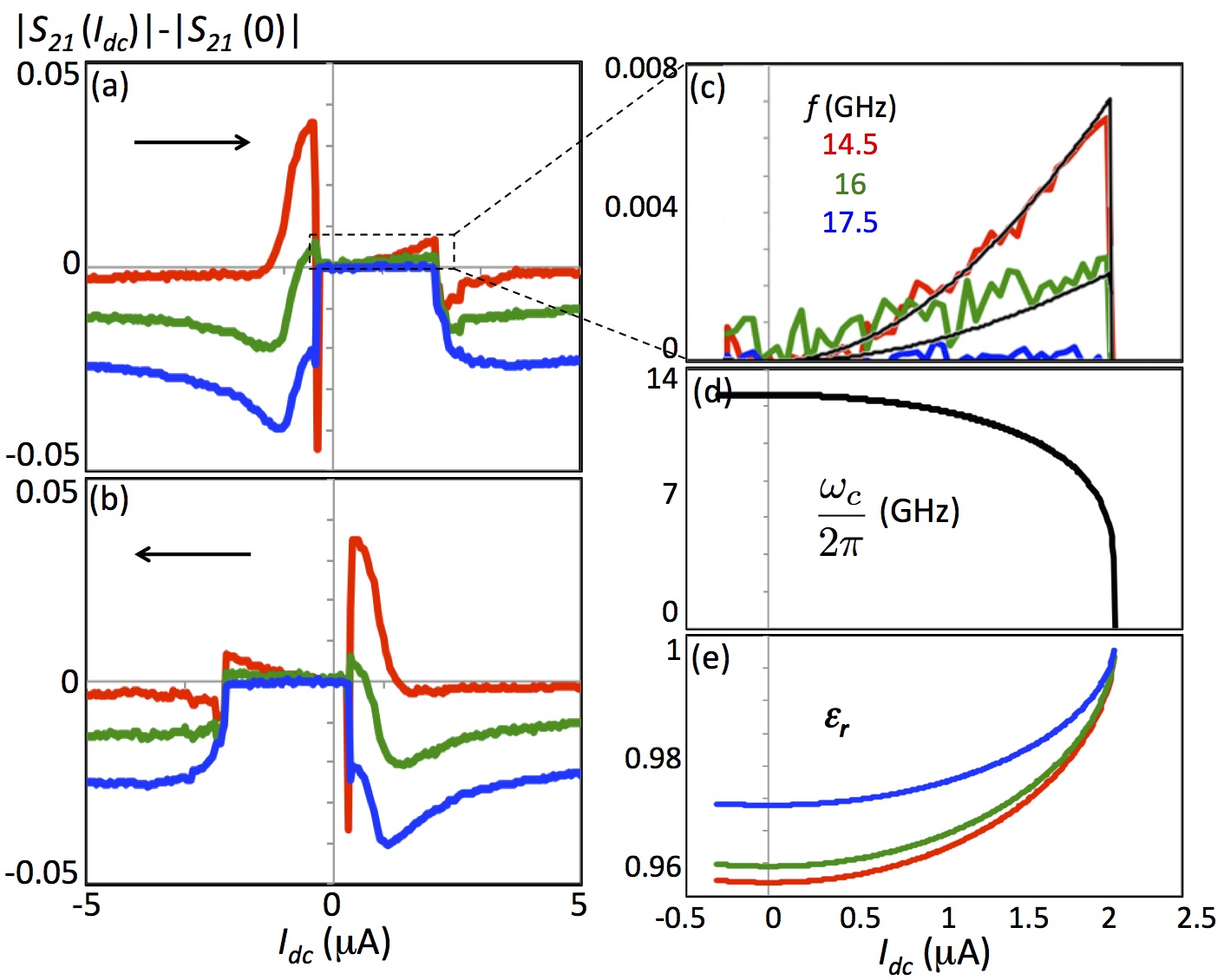}
\caption{Measured transmission magnitude (normalized by $|S_{21}(I_{dc}=0)|$) as a function of increasing (a) and decreasing (b) dc current for three frequencies (all above the cutoff frequency of the wire array in the waveguide) in the limit $I_{rf}\ll I_c$ at $T=4.5$ K. (c) Enlargement of the low dc current region in (a). Black lines are calculated from the model. (d) Simulated waveguide cutoff frequency and (e) simulated effective relative permittivity presented by the JJ-loaded wire array as a function of dc current.}
\label{wirefreq}
\end{figure}

The measured dc current dependence of transmission through a waveguide with a single layer of eight wires for various frequencies is shown in Fig. \ref{wirefreq} (a,b) for different sweep directions of the dc current. The quantity plotted is the change in transmission magnitude $|S_{21}(I_{dc})|$ normalized by $|S_{21}(I_{dc}=0)|$ associated with the presence of a dc bias current through the junctions.  We see that this change in transmission is dependent on both dc current and frequency of the microwave signal.  There is a relatively flat region around zero current (enlarged in Fig. \ref{wirefreq} (c)), but when the dc current approaches the critical current of the array ($I_{c, array} = 8 \times 0.25 \mu A = 2.0 \mu A$) there is a precipitous drop in transmission as $\delta(t)$ begins to develop phase slips and the junctions switch into a more dissipative state.  It is clear from comparison of Fig. \ref{wirefreq} (a) and (b) that the system is hysteretic; the state (dissipative or non-dissipative) of the system depends on the direction of the current sweep.

The model outlined above is only valid for dc current less than the critical current. The model is fit to the data in Fig. \ref{wirefreq}(c) using $F$ as the only fitting parameter.  The model predicts that the cutoff frequency shifts to a lower frequency as the dc current increases, Fig. \ref{wirefreq} (d), and this shift causes the increase in transmission at fixed frequency seen in Fig. \ref{wirefreq} (a and c). The change in transmission is smaller at higher frequencies simply because those frequencies are further above the cutoff frequency. The measured transmission shows good agreement with the model's predictions within its range of applicability.

The shifting cutoff frequency means the effective permittivity of the superconducting metamaterial can be tuned with dc current, as shown in Fig. \ref{wirefreq} (e). The effect is of limited size because there is only one layer of wires with a filling fraction $F=0.05$. Increasing the filling fraction would increase the extent of effective dielectric tuning.

\begin{figure}[h]
\includegraphics*[width=75mm]{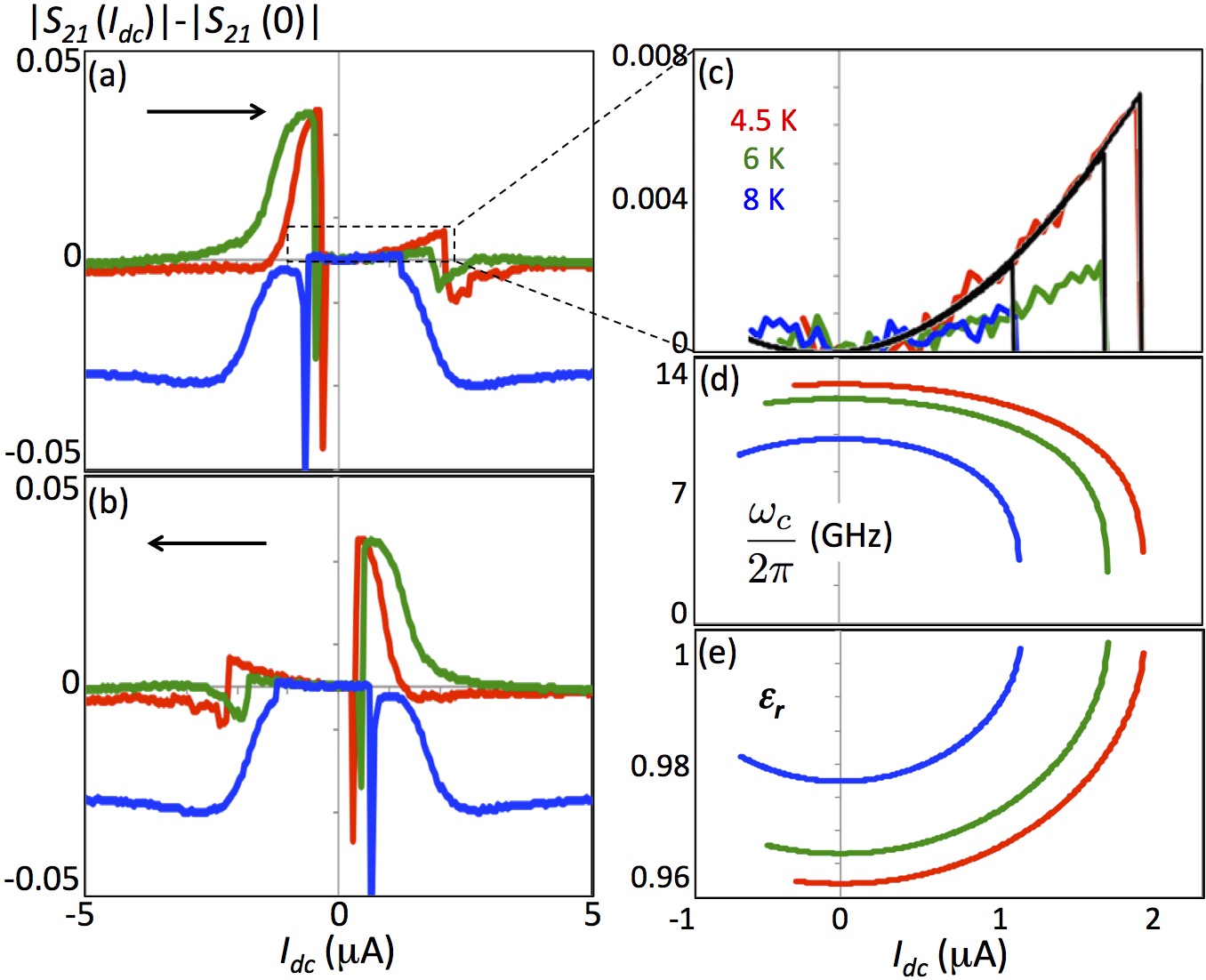}
\caption{(a) Measured transmission magnitude (normalized by $|S_{21}(I_{dc}=0)|$) as a function of increasing (b) and decreasing dc current for three temperatures at 14.5 GHz in the limit $I_{rf}\ll I_c$. (c) Enlargement of the low dc current region in (a). Black lines are calculated from the model. (d) Simulated waveguide cutoff frequency and (e) simulated effective relative permittivity as a function of dc current.}
\label{wiretemp}
\end{figure}

Tuning of the effective permittivity with temperature was also investigated.  Figure \ref{wiretemp} (a,b) shows that the transmission magnitude $|S_{21}(I_{dc})|$ normalized by $|S_{21}(I_{dc}=0)|$ drops (and the junctions transition to the more dissipative state) at lower currents for higher temperatures \textit{i.e.} increasing the temperature lowers the critical current of the junctions. The retrapping current $I_r$, the current at which the junctions switch from the dissipative to the non-dissipative state, increases with temperature. Consequently, the low dc current region (where the model is valid) becomes narrower and less hysteretic.

The model is fit to the data in Fig. \ref{wiretemp}(c) using the measured $I_c(T)$, with $F$ as the only fitting parameter. The model predicts that increasing the temperature decreases the cutoff frequency (Fig. \ref{wiretemp} (d)) increasing the transmission, which agrees with the data, Fig. \ref{wiretemp} (c). Changing temperature is another way to tune the effective permittivity of the superconducting metamaterial, Fig. \ref{wiretemp} (e). However, increased temperature also brings additional dissipation (not modeled here), leading to some degradation of the transmission response.  

As seen in rf SQUID metamaterials, at higher values of the rf and dc currents the junctions enter new regimes of bi-stability \cite{Zhang2015}, multi-stability \cite{Jung2014, TrepPRE2017} and nonlinearity \cite{DaimengIMD2016, Kiselev2019}.  The properties of the JJ array for dc currents exceeding the critical current are also influenced by the increased dissipation in the finite-voltage state of the junctions and will be addressed in future work.

Finally, we note that detecting plasmonic excitations on tunable superconducting wire metamaterials may enable dark matter searches using axion plasma haloscopes \cite{Lawson2019}.

\section{Conclusions}
The Josephson junction-loaded superconducting metamaterial wire array behaves as expected in the low dc current and low input rf power regime it was designed to operate in. The measured tuning of transmission indicates the tuning of the metamaterial cutoff frequency and permittivity predicted by the model. We have successfully demonstrated a tunable dielectric metamaterial based on low-dissipation Josephson inductance tuning.

\begin{acknowledgments}
This work is supported by the NSF-GOALI and OISE programs through grant $\#$ECCS-1158644, the U.S. Department of Energy, Office of Basic Energy Sciences, Division of Materials Sciences and Engineering under Award DESC0018788, and the Center for Nanophysics and Advanced Materials (CNAM).  The fabrication of the Nb-AlOx-Nb circuits was carried out at the Kotel’nikov IREE using USU 352529 facilities and was supported by the Russian Science Foundation (Project No. 19-19-00618).  The authors acknowledge the University of Maryland supercomputing resources made available for conducting the research reported in this paper. We thank M. Radparvar and G. Prokopenko for helpful suggestions and N. Lazarides, G. P. Tsironis, Huan-Kuang Wu, Eileen Stauffer, Min Zhou, P. Jung and S. Butz for productive discussions. We also thank H. J. Paik and M. V. Moody for use of the pulsed tube refrigerator.
\end{acknowledgments}

\bibliography{Bib}
\end{document}